\documentclass[preprint2]{aastex}

\shorttitle{Discovery of a T dwarf via methane imaging}
\shortauthors{Ellis et al.}

\newcommand\lsim{\mathrel{\hbox{\rlap{\hbox{\lower4pt\hbox{$\sim$}}}\hbox{$<$}}}}

\begin{document}

\title{The 2MASS Wide-Field T Dwarf Search.  V.  Discovery of a T Dwarf via Methane Imaging} 

\author{S.~C. Ellis and C.~G. Tinney}
\affil{Anglo-Australian Observatory, P.O. Box 296, Epping, NSW 2121, Australia\\sce@aao.gov.au\\cgt@aao.gov.au}

\author{Adam~J. Burgasser\altaffilmark{1}}
\affil{Department of Astrophysics, Division of Physical Sciences, American Museum of Natural History, Central Park West at 79th Street, New York, NY 10024, USA\\adam@amnh.org}

\author{J.~Davy Kirkpatrick}
\affil{Infrared Processing \& Analysis Center, Caltech, Pasadena, CA 91125, USA\\davy@ipac.caltech.edu}

\and

\author{Michael W.~McElwain}
\affil{UCLA, 8371 Mathematical Sciences, CA 90095, USA\\mcelwain@astro.ucla.edu}

\altaffiltext{1}{Spitzer Fellow}


\begin{abstract}
We present the discovery of a T dwarf, 2M2151-4853, via differential imaging through methane filters.  The filters are designed to highlight the strong absorption in the H band, due to methane found in the atmospheres of T dwarfs, and provide a very efficient means of searching for them.  Subsequent J and H band spectroscopy confirms 2M2151-4853 as a T dwarf of type T4.5.  It has an estimated spectrophotometric distance of 18$\pm$3 pc, and an estimated tangential velocity of $v_{{\rm t}}=50 \pm 10$ km s$^{-1}$.
\end{abstract}


\section{INTRODUCTION}

T dwarfs are the coolest ($T\lsim 1500$K, \citealt{vrb04}; \citealt{gol04}) and least luminous brown dwarfs observed.  With atmospheres rich in molecular gases and condensate clouds (e.g.\ \citealt{ack01}),  these sources represent a class of object intermediate between giant planets and low mass stars. 

In the cool atmospheres of T dwarfs, methane and water vapor are formed in abundance.  These give rise to broad absorption features in the near-infrared, which distinguish T dwarfs from hotter brown dwarfs in which the production of atmospheric methane is prohibited by collisional dissociation (\citealt{nol00}).  These prominent CH$_{4}$ and H$_{2}$O features have been used to construct schemes of spectral classification for these objects (\citealt{burg02}; \citealt{geb02}).  Because of their low photospheric temperatures, an understanding the atmospheres of T dwarfs is likely to yield important clues about the nature of giant planetary atmospheres.

A statistically complete census of T dwarfs is highly desirable in an effort to extend our understanding of the initial mass function to lower masses, but
cataloguing T dwarfs can be a time consuming process.  With faint intrinsic magnitudes and near-infrared colors similar to most main sequence stars, T dwarf candidates can be buried in an overwhelming number of background sources.  Traditional spectroscopic follow-up of large samples is extremely time consuming and has typically had low rates of success. 
\citet{tin05} describe a new method for simplifying the search process, based on imaging through CH$_{4}$ filters.  Here we describe a new T dwarf discovered by this method.

Section~\ref{sec:sample} summarizes the project and the methane imaging process.  Section~\ref{sec:results} describes the observations and empirical properties of the new T dwarf 2MASS J21513839-4853542, hereafter 2M2151-4853.  Conclusions are given in Section~\ref{sec:conc}.

\section{THE 2MASS WIDE-FIELD T DWARF SEARCH AND METHANE IMAGING}
\label{sec:sample}

The 2MASS Wide-Field T Dwarf Search is an attempt to conduct a census of the T dwarf population over most of the sky.  The selection process is described in detail by \citet{burg03}, and is summarized here.

The initial sample has been culled from the 2MASS All Sky Working Database (\citealt{skr97}; \citealt{cut03}), via color and magnitude selection.  Areas within a galactic latitude $|b|<15^{\circ}$ have been excluded to avoid regions where source confusion will be problematic.  Likewise the Magellanic clouds, and other dense source regions have been omitted.  To facilitate follow up observations, regions within 2$^{\circ}$ of the equatorial poles were also omitted.  In total the search covers 74\% of the sky.

From this sample objects were selected with $J-H \le 0.3$ or $H-K_{{\rm S}} \le 0$, $J \le 16$, and no counterpart in the USNO-A2.0 catalogue (\citealt{mon98}), effectively $R-J>4$.  This range of color is sufficiently broad to accommodate warmer T dwarfs, as early as type T1 (\citealt{tin05}).

This selection process yields over 250,000 targets.  Since it is expected that there will only be 20-30 T dwarfs in this sample (\citealt{burg03}), further discrimination of the targets has been made on the basis of visual classification by AJB and MWM, leaving $\sim 1500$ targets.  Targets from this list are being observed as part of a methane imaging program.
 

Our methane imaging program of 2MASS T dwarf candidates is described in full by \citet{tin05}.  Images were taken through the CH$_{4}$s and CH$_{4}$l filters with the infrared camera and spectrograph (IRIS2) on the 3.9m Anglo-Australian Telescope, Siding Spring, Australia.

The CH$_{4}$l filter covers the wavelength range of the strong methane absorption features in the infrared H band, which define the spectral type T dwarfs (\citealt{geb02}).  Meanwhile the CH$_{4}$s filter samples a wavelength range outside the methane absorption (though within the range of absorption by other molecules such as H$_{2}$O).

The methane observations were calibrated via a series of extensive observations described in \citet{tin05}.  Stars of known spectral standard were observed, and then calibrated according to observations of A, F and G stars which show virtually no color variation in CH$_{4}$s$-$CH$_{4}$l, and K and M stars for which the methane colors have been empirically calibrated.  

\section{OBSERVATIONS}
\label{sec:results}

\subsection{Methane Imaging}

\begin{figure*}
\centering \includegraphics[angle=0,scale=0.5]{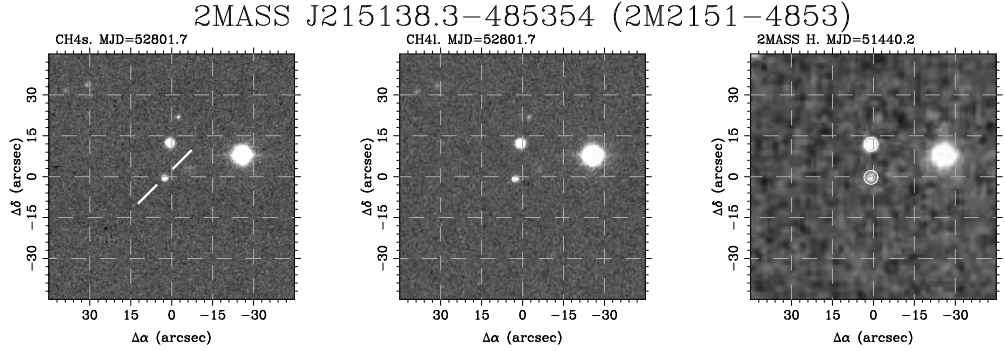}
\caption{Finding chart for 2M2151-4853.  The images are 1.5 arcminutes on a side, and are centered on the 2MASS coordinates, 21$^{{\rm h}}$51$^{{\rm m}}$38.39$^{{\rm s}}$, -48$^{\circ}$53$'$54.2$''$, and oriented with North up and East toward the left.  The slight offset from the 2MASS coordinates illustrates the substantial proper motion of 2M2151-4853 of 0.57$\pm0.07$ arcseconds year$^{-1}$, see Section~\ref{sec:motion}.}
\label{fig:find}
\end{figure*}

\begin{figure*}
\centering \includegraphics[scale=0.5,angle=270]{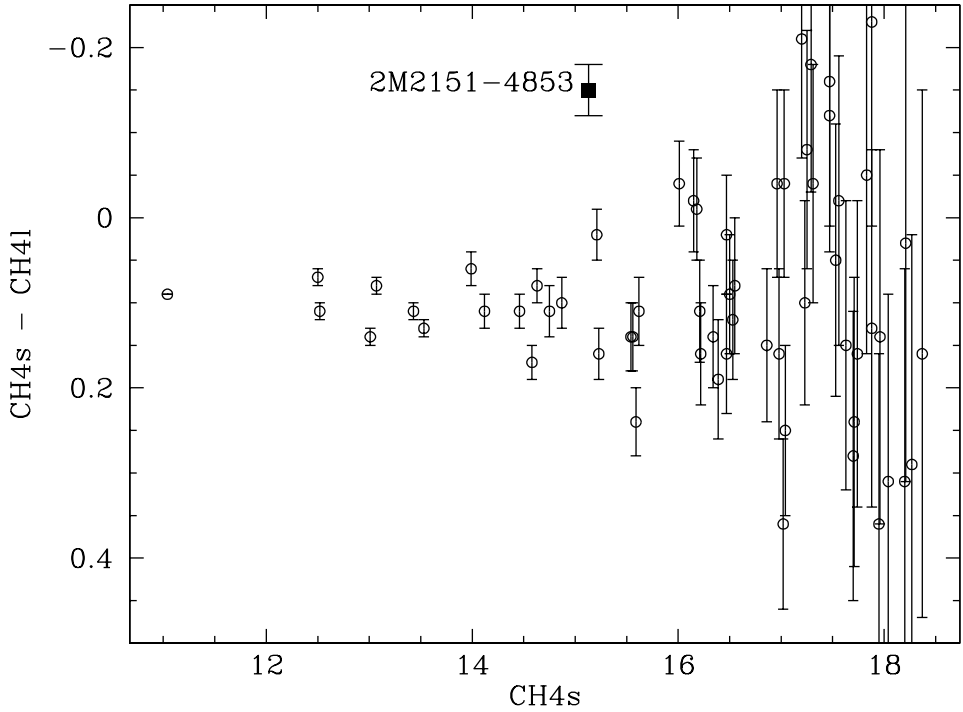}
\caption{CH$_{4}$s$-$CH$_{4}$l vs. CH$_{4}$s color-magnitude diagram.  2M2151-4853 is plotted as the filled square and is obvious at CH$_{4}$s=15.13  by its excess methane color.}
\label{fig:colmag}
\end{figure*}

Methane observations of 2M2151-4853, as described above, were made on 2003 June 11 (UT), in patchy cloud (which appears not to have affected our imaging) and 1.1 arcsec seeing.  Exposures of 30 seconds in each of the CH$_{4}$s and CH$_{4}$l filters were made.  Images were dark subtracted and flat fielded using standard procedures with the pipeline reduction software, {\sc orac-dr}.  The flat fields were made by median combining jittered images of the targets.  Finding charts extracted from these images and the 2MASS H band image are shown in Figure~\ref{fig:find}.  {\sc SExtractor} (\citealt{ber96}) was used to measure magnitudes in fixed apertures of $\approx 2.5$ arcsec diameter.  The differential photometry was calibrated using the procedure described in \citet{tin05} using 2MASS sources in the field, to produce Figure~\ref{fig:colmag}. 

\begin{figure*}
\centering \includegraphics[angle=270,scale=0.5]{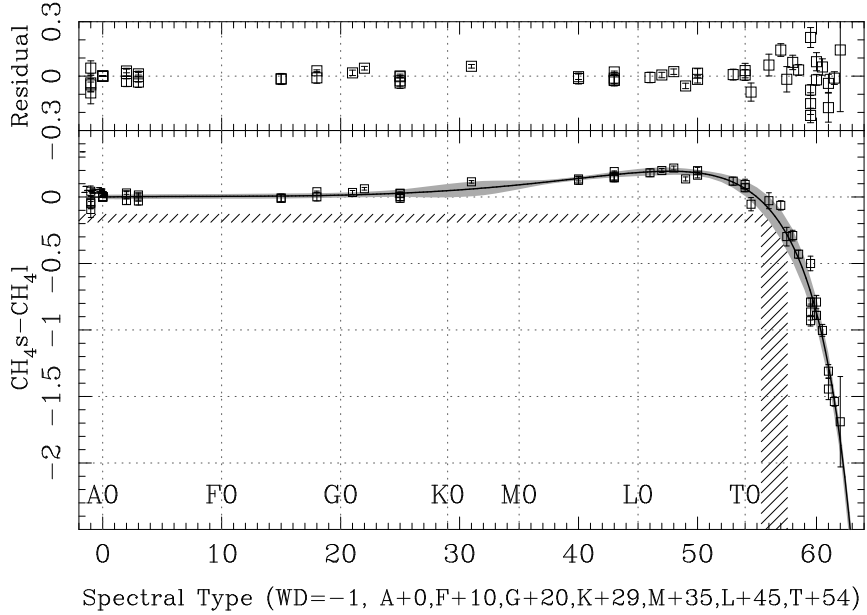}
\caption{Methane color as a function of spectral-type.  The shaded region around the curve represents the uncertainties derived from photon counting statistics, aperture correction uncertainties and photometric calibration uncertainties.  The horizontal and vertical hatched region represents the methane color and uncertainty of 2M2151-4853 and the corresponding spectral type.  See \protect\citet{tin05} for full details.}
\label{fig:methane_id}
\end{figure*}

A plot of CH$_{4}$s$-$CH$_{4}$l against CH$_{4}$s is shown in Figure~\ref{fig:colmag} with 2M2151-4853 clearly visible above the other objects in the field.    A comparison of the methane color of 2M2151-4853 with the calibrated methane colors of A stars through to T dwarfs is shown in Figure~\ref{fig:methane_id}, taken from \citet{tin05}.  The color of CH$_{4}$s$-$CH$_{4}$l=$-0.15 \pm 0.03$ suggests that 2M2151-4853 is of type T2.5$\pm 1$.  Figure~\ref{fig:methane_id} highlights the power of the methane imaging methodology.  Short images of the target field are made through the CH$_{4}$ filters.  Any T dwarfs present in the field will be immediately obvious in a plot of CH$_{4}$s$-$CH$_{4}$l, standing out prominently from the locus of points made up from the hotter stars.  Furthermore, a first estimate of spectral type within the T dwarf class can be made from the methane color.  Thus only T dwarf candidates with unusual methane colors ever need to be followed up with spectroscopic observations.  

\subsection{Spectroscopy}

Hs band spectroscopy of  2M2151-4853 was obtained using IRIS2 on 2004 June 23 (UT).  The seeing was $\sim 1.1$ arcseconds, and there was patchy cloud cover.  This spectrum was taken with  very low signal to noise, sufficient to identify 2M2151-4853 as a T dwarf, but not to classify its spectral type.  Higher signal to noise spectra spectra were then obtained on 2004 August 22 (UT) under good conditions (seeing $\sim 1.2$ arcseconds and clear sky) in the Hs and Jl bands. The total exposure was 15 minutes in each band.  The spectra were flat-fielded using dome flats created from a lamp-on$-$lamp-off sequence.  The wavelength calibration was achieved through imaging of a Xe arc lamp.  The flattened spectra were divided by a normalized spectrum of a G dwarf telluric standard star to remove the effect of absorption bands.

\begin{figure*}
\begin{minipage}[c]{0.5\textwidth}
\centering \includegraphics[scale=0.3,angle=270]{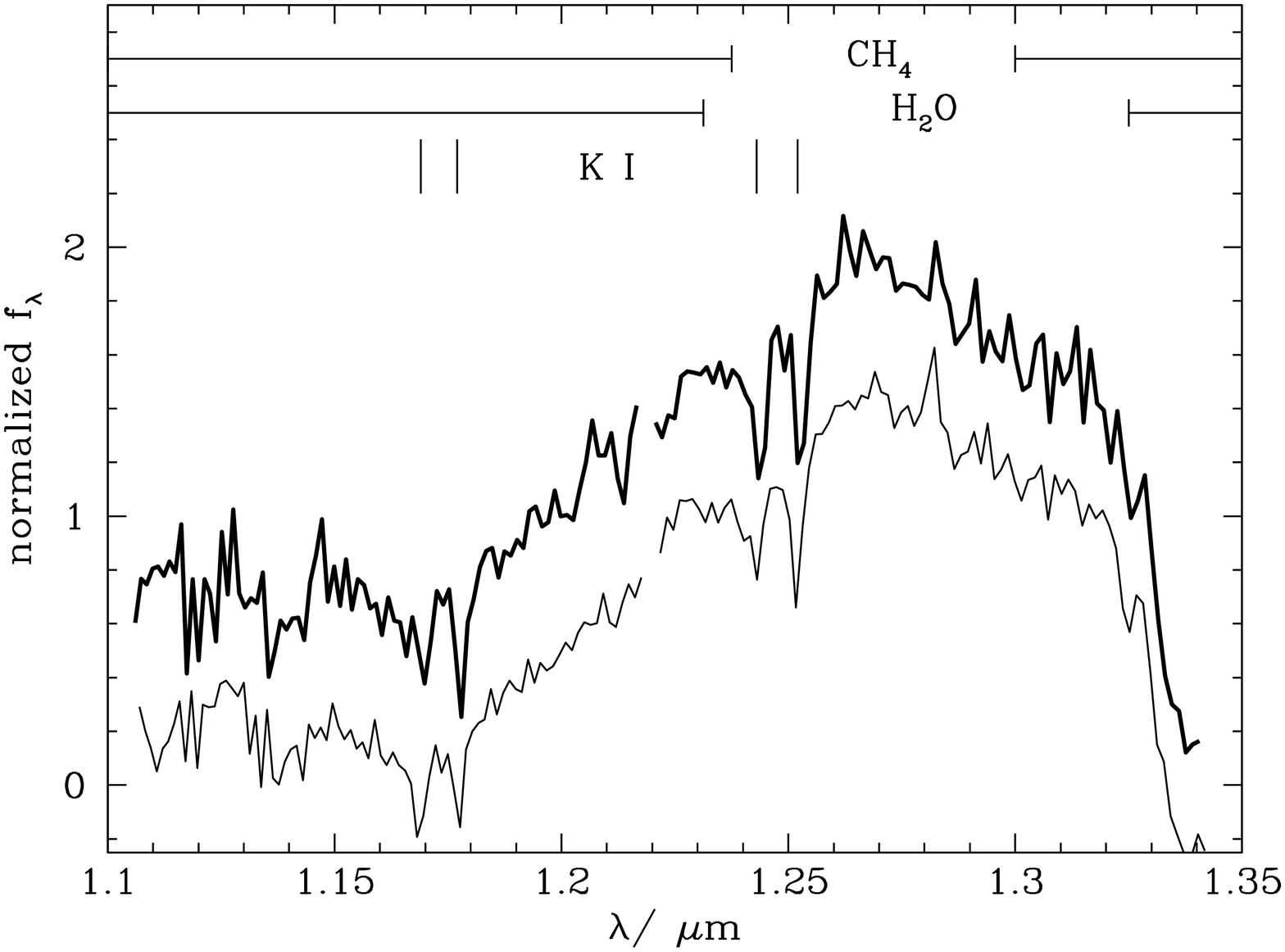}
\end{minipage}%
\begin{minipage}[c]{0.5\textwidth}
\centering \includegraphics[scale=0.3,angle=270]{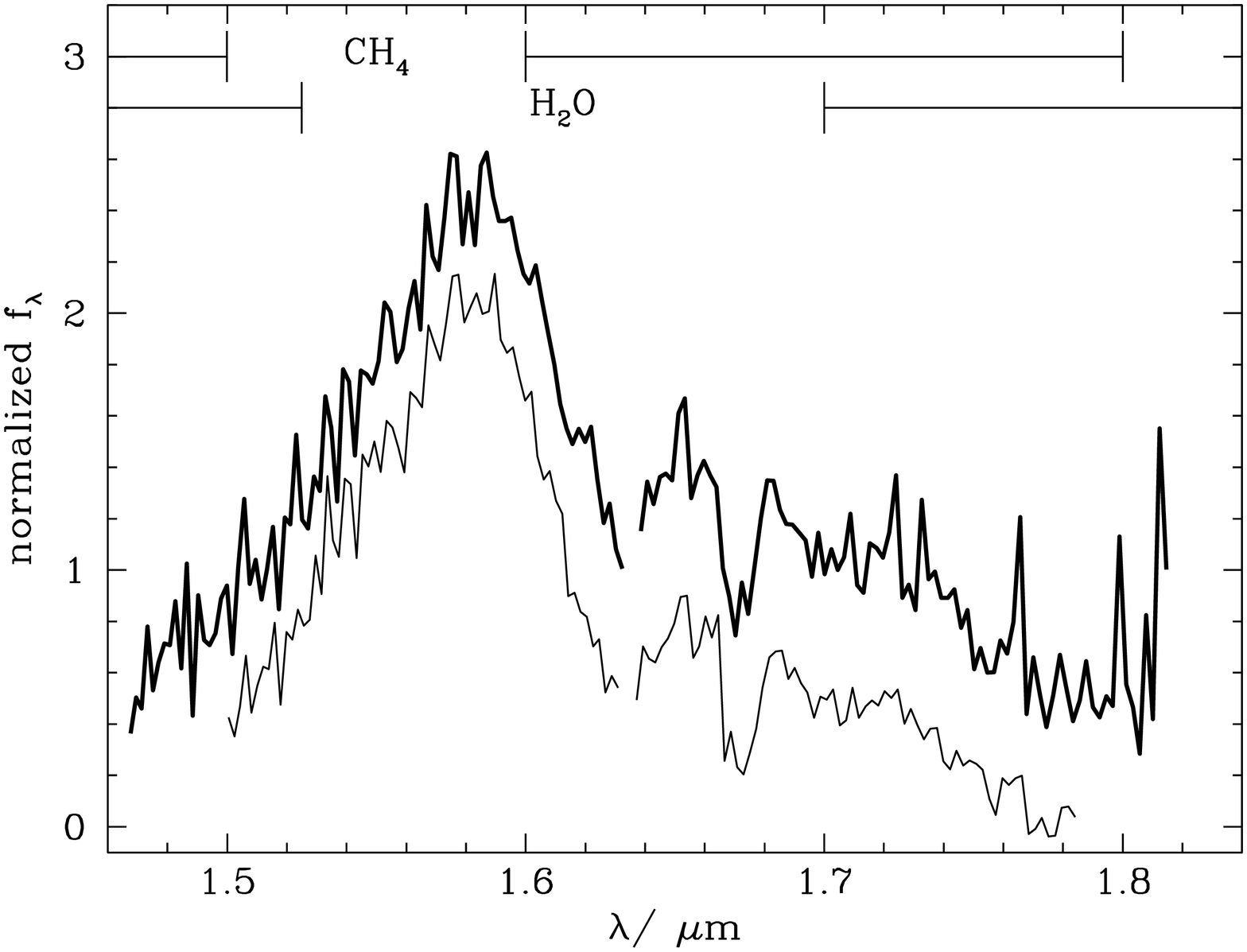}
\label{fig:spect2}
\end{minipage}
\caption{The Jl (left panel) and Hs (right panel) band spectra of 2M2151-4853 are shown by the upper bold lines.  The lower spectra are those of the T4.5 dwarf, 2M0559-1404, and are offset by a constant of 0.5 for clarity.  The Jl band spectra have been normalized at $\lambda=1.2\mu$m and the Hs band spectra at $\lambda=1.7\mu$m.   The H$_2$O and CH$_4$ absorption bands are indicated along with the K{\sc i} absorption lines.}
\label{fig:spect}
\end{figure*}

The final spectra, binned by a factor of six to improve signal to noise, are shown in Figure~\ref{fig:spect}, along with the spectrum of the T4.5 dwarf, 2MASS J05591914-1404488, hereafter 2M0559-1404 (\citealt{burg00}; \citealt{tin05}).   The distinctive absorption features due to H$_{2}$O and CH$_{4}$ are clearly visible, and are identified at the top of each figure, along with the K{\sc i} absorption lines in the Jl band.

Spectral classification schemes have been devised by \citet{burg02} and \citet{geb02}.  
We have used the spectral indices of both schemes within the wavelength range of our spectra, to check the classification based on the methane colors described above.

Both schemes define spectral indices based on distinctive H$_{2}$O and CH$_{4}$ features in T dwarf spectra.  \citet{burg02} define the diagnostics given in equation~\ref{eqn:burgindex}, based on the mean flux within the regions of interest, and those of a neighboring region of the spectrum outside the absorption feature, and calculating their ratio.  For the case of 2M2151-4853 we have,

\begin{eqnarray}
\label{eqn:burgindex}
{\rm H_{2}O}-A=\frac{<F_{1.12-1.17}>}{<F_{1.25-1.28}>}=0.37\\ \nonumber
{\rm H_{2}O}-B=\frac{<F_{1.505-1.525}>}{<F_{1.575-1.595}>}=0.46\\ \nonumber
{\rm CH_{4}}-A=\frac{<F_{1.295-1.325}>}{<F_{1.25-1.28}>}=0.81\\ \nonumber
{\rm CH_{4}}-B=\frac{<F_{1.640-1.700}>}{<F_{1.575-1.595}>}=0.49,\nonumber
\end{eqnarray}

where the subscripts denote the wavebands of interest in $\mu$m.  \citet{burg02} list index values for T dwarf standards in their table~10.  A comparison with that table shows that 2M2151-4853 lies between type T3 and T5 for all the diagnostics (there is no type T4 standard listed).

\citet{geb02} define similar indices based on the integrated flux from diagnostic regions of T dwarf spectra.  Computing their indices for 2M2151-4853 we find,

\begin{eqnarray}
\label{eqn:gebindex}
{\rm CH_{4} index}=\frac{\int_{1.56}^{1.60} f_{\lambda}{\rm d}\lambda}{\int_{1.635}^{1.675} f_{\lambda}{\rm d}\lambda} =1.82\\ \nonumber
{\rm H_{2}O index}=\frac{\int_{1.26}^{1.29} f_{\lambda}{\rm d}\lambda}{\int_{1.13}^{1.16} f_{\lambda}{\rm d}\lambda} =2.66,\nonumber
\end{eqnarray}

where the limits of the integrals are given in $\mu$m.  \citet{geb02} present a look-up table of indices for different spectral types. Comparing to this (table 5 in their paper), we find that the CH$_{4}$ and H$_{2}$O indices indicate that 2M2151-4853 is of type T5 and T4 respectively.

A hybrid system between the schemes of \citet{burg02} and \citet{geb02} is in development, utilizing a unified set of fundamental T dwarfs and spectroscopic indices (\citealt{burg05}).  We compared a subset of the revised spectral indices for this scheme (presented in \citealt{burg03b}) between the spectrum of 2M2151-4853 and those of a set of T dwarf comparison sources (\citealt{tin05}).  These indices yield a mean spectral type of T4.5 for 2M2151-4853, consistent with the above indices.  This classification is further verified by the similarity of J and H band spectra with those of 2M0559-1404 (Figure~\ref{fig:spect}), itself a T4.5 dwarf (\citealt{geb02}).

\subsection{Spectrophotometric Distance}

The 2MASS photometry available for 2M2151-4853 has been used to derive an estimate of distance.  We have compared the 2MASS magnitudes to the spectral type--absolute magnitude relations given by \citet{tin03} and \citet{vrb04}.  In order to make the comparison with \citet{vrb04} the 2MASS magnitudes were first transformed to the CIT magnitude system using the relations of \citet{ste04}.  The results are given in Table~\ref{tab:mags}.

\begin{table*}
\begin{center}
\caption{Photometry and derived distances}
\label{tab:mags}
\begin{tabular}{llll}
\hline \hline
& J & H  & K/Ks \\ \tableline
Apparent magnitude 2M2151(2MASS) &$15.73 \pm 0.08$ & $15.17 \pm 0.10$ & $15.43 \pm 0.18$ \\
Apparent magnitude 2M2151(CIT)$^{a}$  & $15.55 \pm 0.08$ & $15.22 \pm 0.10$ & $15.57 \pm 0.19$ \\
Expected absolute magnitude (2MASS, \protect\citealt{tin03}) & $14.0\pm0.4$ & & $14.0\pm 0.4$\\ 
Expected absolute magnitude (CIT, \protect\citealt{vrb04}) & 14.5 & 14.3 & 14.2\\
Derived distance /pc (\protect\citet{tin03}) & 23 & & 19\\ 
Derived distance /pc (\protect\citealt{vrb04}) & 16 & 15 & 19 \\ \hline
\end{tabular}
\tablenotetext{a}{Calculated from the near-infrared color transformations of \citet{ste04}}
\end{center}
\end{table*}

The distances calculated from the relations of \citet{tin03} and \citet{vrb04} are $21\pm 2$pc and $17\pm 2$pc respectively.  The average distance using all relations is 18 $\pm$ 3 pc assuming it is single.

\subsection{Proper Motion}
\label{sec:motion}

2M2151-4853 has an appreciable proper motion measured at $0.57 \pm 0.07$ arcseconds year$^{-1}$ with a position angle of 113$\pm$3 degrees, based on the difference in position of the 2MASS images and the methane images.  Using the distance estimate derived above this translates to a tangential velocity of $v_{{\rm t}}=50 \pm 10 $ km s$^{-1}$.

\section{CONCLUSIONS}
\label{sec:conc}

The discovery of 2M2151-4853 further emphasizes the power and efficiency of methane imaging in searching for T dwarfs.  This method has been used to great success in the 2MASS wide-field T dwarf search, as demonstrated in \citet{tin05}.


We have used the spectral classification schemes of \citet{burg02}, \citet{geb02} and \citet{burg05} to classify 2M2151-4853.  All schemes are consistent with 2M2151-4853 being of type T4.5.  This is confirmed in a comparison of the spectra of 2M2151-4853 with those of the  T4.5 dwarf, 2M0559-0414.  Spectrophotometric distances indicate it is at a distance of 18$\pm$3 pc and its proper motion implies a tangential velocity of $v_{{\rm t}}=50 \pm 10$km s$^{-1}$, consistent with disk population kinematics.  The properties of 2M2151-4853 are summarized in Table~\ref{tab:summ}.

A total of 11 T dwarfs have now been identified in the 2MASS sample via methane imaging; however, our candidate sample is not yet exhausted.  Further discoveries and a final accounting of the search will be provided in a forthcoming paper.

\begin{table*}
\begin{center}
\caption{Summary of the Properties of 2M2151-4853}
\label{tab:summ}
\begin{tabular}{ll}
\hline \hline
R.A. at 2003 June 11 (UT J2000.0) & $21^{{\rm h}}51^{{\rm m}}38^{{\rm s}}.6$ \\
Dec. at 2003 June 11 (UT J2000.0) & $-48^{\circ}53'55''$\\
2MASS J mag & 15.73$\pm0.07$ \\
2MASS H mag & 15.17$\pm0.10$ \\
2MASS Ks mag & 15.43$\pm0.18$ \\
CH4s mag & 15.13$\pm0.02$ \\
CH4l mag & 15.28$\pm0.02$ \\
Proper motion & 0.57$\pm 0.07$ arcsec yr$^{-1}$ \\
Position angle & 113 $\pm 3$ degrees\\
Distance & 18$\pm 3$ pc \\
Tangential velocity & 50$\pm 10$ km s$^{-1}$ \\ \hline
\end{tabular}
\end{center}
\end{table*}

\section*{Acknowledgments}
The authors wish to gratefully acknowledge the Joint Astronomy Centre, Hawaii, for making their {\sc orac-dr} code available, and for assisting the AAO in implementing it for IRIS2.
SCE wishes to acknowledge PPARC support whilst working on this paper.  AJB acknowledges support by NASA through the SIRTF Fellowship Program.

This publication makes use of data from the Two Micron All Sky Survey which is a joint project of the University of Massachusetts and the Infrared Processing and Analysis Center, funded by NASA and the NSF.

2MASS data were obtained from the NASA/IPAC Infrared Science Archive, which is operated by the Jet Propulsion Laboratory, California Institue of Technology, under contract with the National Aeronautics and Space Administration.


\end{document}